\documentclass[12pt,preprint]{aastex}

\shorttitle{Gas-phase CO and mixing in PP disks}
\shortauthors{Semenov et al.}

\begin{document}

\title{Gas-phase CO in protoplanetary disks: A challenge for turbulent mixing}

\author{D. Semenov}
\affil{Max--Planck--Institut f\"ur Astronomie, K\"onigstuhl 17, D--69117 Heidelberg, Germany}
\email{semenov@mpia.de}

\author{D. Wiebe}
\affil{Institute of Astronomy of the RAS, Pyatnitskaya St. 48, 119017 Moscow, Russia}
\email{dwiebe@inasan.ru}

\and

\author{Th. Henning}
\affil{Max--Planck--Institut f\"ur Astronomie, K\"onigstuhl 17, D--69117 Heidelberg, Germany}
\email{henning@mpia.de}

\begin{abstract}
This is the first paper in a series where we study the influence
of turbulent diffusion and advective transport on the chemical
evolution of protoplanetary disks, using a 2D flared disk model
and a 2D mixing gas-grain chemical code with surface
reactions. A first interesting result concerns the abundance of
gas-phase CO in the outer regions of protoplanetary disks. In this Letter
we argue that the gas-phase CO concentration in the disk regions, where the temperature is
lower than $\sim25$\,K, can be significantly enhanced due to the combined effect
of vertical and radial mixing. This finding has a potential implication
for the current
observational data on the DM~Tau disk chemistry.
\end{abstract}

\keywords{accretion, accretion disks --- astrochemistry ---
circumstellar matter --- molecular processes --- planetary
systems: protoplanetary disks --- turbulence}

\section{Introduction}
The angular momentum redistribution and associated global mass transport
phenomena (advection, turbulent diffusion) constitute an inherent
factor of protoplanetary disk evolution
\citep[e.g.,][]{MV1984,CP1988,JK05}. There are many observational facts
indicative of both large-scale and small-scale radial
mixing, e.g, the presence of crystalline dust in comets
\citep{bockelee,wooden} and protoplanetary disks
\citep{gail2001,vanBoekelea04}. Radial mixing has also been invoked to
explain the
distribution of water ice in the Solar Nebula \citep{cyrp1998}, deuteration in the Solar Nebula \citep{hersant2001}, and isotopic abundances in meteorites \citep{boss2004}.

Because of the complexity of the transport problem, disk chemistry has been
often treated without considering dynamics or only with inward advection
\citep[e.g.,][]{bauerea97,willacyea98,aikawaea99,mk2002}. By
now, there are only three papers where the turbulent
transport in protoplanetary disks is treated in conjunction with
the time-dependent chemistry. \citet{ilgnerea04} have considered
vertical turbulent diffusion and radial advection in the inner part of a
protoplanetary disk (in the 1+1D approximation) to study the
effect of mixing on sulfur chemistry. \citet{ilgner06b} studied the
impact of vertical diffusive turbulent transport on the fractional ionization in the inner part of a
protoplanetary disk. \citet{willacy2006} considered the influence
of vertical turbulent diffusion on the chemical evolution of the outer disk.
They have shown that vertical diffusion greatly affects the column densities
of many species and may have an effect detectable in millimeter and sub-millimeter
observations.

There are molecular line observations of the outer parts of protoplanetary disks
that cannot easily be explained with chemical models for static disks. One of
the puzzles is the abundant gas-phase CO in the DM~Tau outer disk,
where the kinetic temperature is lower than $\sim20$\,K
\citep{dartois}. In the absence of effective desorption
mechanisms, CO molecules are supposed to be completely frozen out
in the cold midplane. \cite{an2006} suggested that the existence of such cold CO
gas can be explained by vertical mixing.
In this Letter we show that the solution to the enhanced gas-phase CO can be found
within the framework of a 2D non-static (dynamical) chemical disk model, even though
the abundance predicted in the current study is still lower than observed. We
demonstrate that in order to assess the
importance of turbulent transport processes, in general, one has to consider
vertical and radial mixing simultaneously.

\section{Disk model}

The disk physical and chemical model will be comprehensively described
elsewhere (Semenov et al., in preparation). Briefly, we adopt
the disk structure (i.e. density and temperature) from a steady-state irradiated disk model developed by \cite{pad1998,pad1999}.
The disk has a radius of 800~AU, an accretion rate
$\dot{M}=10^{-8}\,M_\sun$\,yr$^{-1}$, a viscosity parameter
$\alpha = 0.01$, and a mass of $M=0.07\,M_\sun$. This value is comparable to the inferred DM~Tau disk mass of 0.05 $M_\odot$ \citep{dmtau97}. The temperature and density
structure of the model disk is shown in Fig.~\ref{disk_struc} (left and middle panels). The vertical coordinate $z'$ in this and other figures is measured in units of the relative height above the disk plane, so that $z'=z/z_{\max}$. The disk semi-thickness $z_{\max}$ is the height at
a constant pressure $P_{\max}= 10^{-10}$ dyn cm$^{-2}$. The ratio $z_{\max}/r$ is 0.9 at 100~AU and 1.5 at 800~AU.

The disk is illuminated by UV radiation from a central star with
an intensity of $G=540\,G_0$ at $r=100$~AU and by interstellar UV
radiation \citep[1D penetration,][]{G,berginea03}. The dust grains
are assumed to be uniform $0.1\,\mu$m spherical particles in the chemical model. This assumption is formally inconsistent with the grain size distribution used in the disk model. However, the relevant parameter for our chemical model is the total dust surface area (per unit volume). It differs from the dust surface area in the disk model by no more than a factor of 2. The self-
and mutual-shielding of CO and H$_2$ is computed as in
\citet{leeea96}. Three other energy sources are cosmic rays,
short-living radionuclides, and stellar X-rays.

The X-ray ionization rate is computed according to
\citet{glassgoldea97b} with parameters for their high depletion
case and the luminosity value $L_{\rm X} \approx
10^{30}$~erg\,cm$^{-2}$\,s$^{-1}$ \citep{glassgoldea05}. A
gas-grain chemical model with surface reactions from
\citet{Semenovea05} is utilized. The gas-phase reaction rates
are taken from the UMIST\,95 database \citep{umist95}, with some rates
updated according to recent measurements. Cosmic-ray induced desorption, X-ray induced desorption, thermal desorption, and photodesorption are taken into account.
Applying the reduction techniques described in \citet{Red1} to a static model with
molecular (TMC1-like) initial abundances, we isolated a network
consisting of 263 species and 1139 reactions. The disk chemical structure, modeled with this network, has been compared with predictions
of the full network for a static model and for models with vertical and radial diffusion. This comparison has shown that the reduced network allows to follow
the evolution of CO, H$_2$CO, and other interesting molecules with an
accuracy of a factor of $\sim 2-3$. In this Letter we utilize this reduced network.

In addition to time-dependent chemistry, we consider radial and
vertical turbulent mixing using the approach of \citet{Xieea95}, in which
turbulence is treated as if it were a diffusion process.
The diffusion coefficient is estimated as
\begin{equation}
D_{\rm turb} =
\epsilon\,\nu=\epsilon\,\alpha\,c_{\rm s}\,H,
\end{equation}
where $\nu$ is the
viscosity, $c_{\rm s}$ is the sound speed, $H$ is the scale
height, and $\epsilon \la 1$ is the efficiency of turbulent
transport \citep[see e.g.][]{SS73,BS02,sh04,JK05}. If the efficiency
parameter is close to unity (as assumed in this study), the typical value of $D_{\rm turb}$
in disks at 100 AU should be about $10^{17}$\,cm$^2$\,s$^{-1}$
(Fig.~\ref{disk_struc}, right panel). Note that the vertical mixing
timescale in this case can be estimated as $t_{\rm mix}
\sim H^2/D_{\rm turb}$, which gives $\sim 10^4$~years for
$r=100$~AU. Advection due to global gas motions is not taken into account.

The equations of chemical kinetics are integrated
simultaneously with the diffusion terms in the Eulerian
description, using a fully implicit 2D integration scheme. 
Radial mixing is in general less of an effect on the disk
chemistry than vertical mixing, while full 2D mixing acts
as a combination of both processes. The latter fact justifies the use
of approximate, fast operator-splitting integration schemes, in
which radial and vertical mixing processes are treated separately, in
addition to time-dependent chemistry \citep[e.g.,][]{KG04}. All the
equations are solved on a non-uniform staggered grid consisting of
30 radial points (from 10 to 800~AU) and 65 vertical points. This resolution was found
to be optimal for mixing problems in a protoplanetary accretion disk. The
evolutionary time span is 5~Myr.

\section{Results and Discussion}
\begin{figure}
\plotone{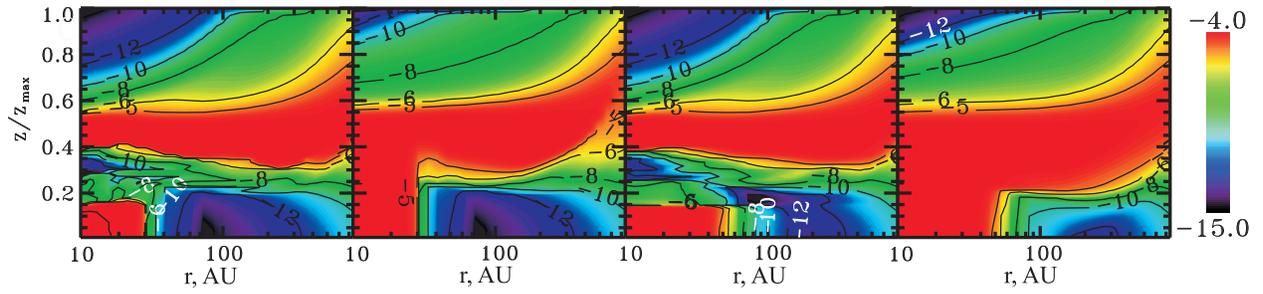}
\caption{The logarithm of relative gas-phase CO abundances (with
respect to the total number of hydrogen nuclei) at 5~Myr for four
chemical models (from left to right): static chemistry,
chemistry with vertical mixing, chemistry with radial mixing, and
chemistry with full 2D mixing. See the electronic edition
of the Journal for a color version of this figure.} \label{abunds}
\end{figure}

Our main set of results consists of four runs (Fig.~\ref{abunds}), with the static chemical model (no transport processes are taken into
account), with vertical mixing only, with radial mixing only, and, finally, with simultaneous mixing in both
directions.

\begin{figure}
\epsscale{0.5}
\plotone{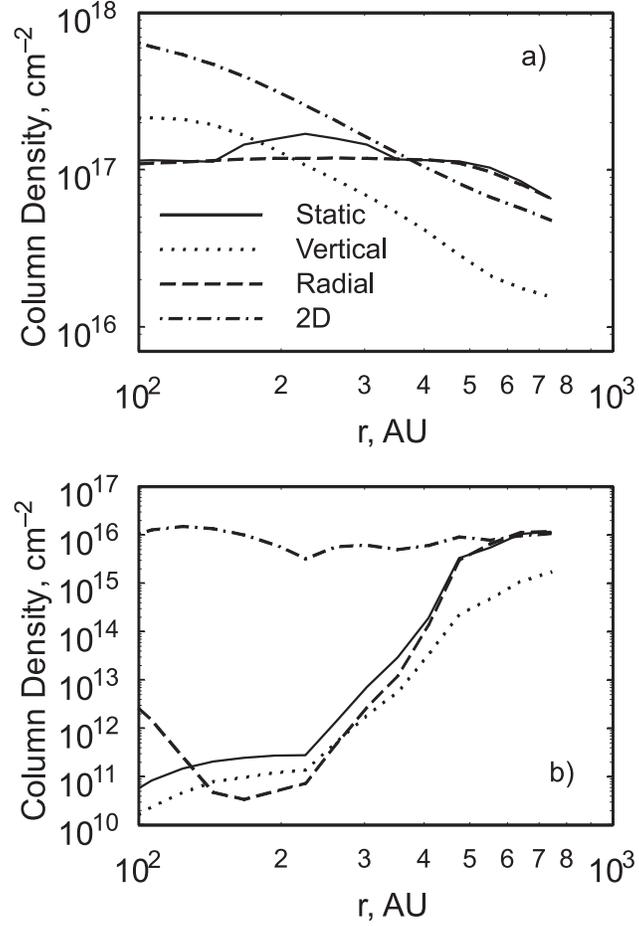}
\caption{CO column densities in the outer disk for all considered models. The total column density is presented on top panel, while on bottom panel only the contribution from the cold region ($T<25$\,K) is shown.} \label{cold}
\end{figure}

The well-known ``sandwich-like'' structure typical of
protoplanetary disks is evident in all plots. The location of
the warm molecular CO layer is defined by photodissociation from above and
freeze-out from below. It is natural to expect that this feature is
especially sensitive to vertical mixing since the abundance
gradient in the vertical direction is very steep. However,
neither of the considered transport processes drastically
changes its abundances in the upper disk, which is further illustrated in Fig.~\ref{cold}.
Total column densities, shown in the top panel of
Fig.~\ref{cold}, differ by less than an order of magnitude in all four models.

This situation is different  in the cold disk midplane, where CO molecules are essentially frozen
out onto dust grains within a few $10^3$ years. After this rapid phase of the CO depletion, slow
surface reactions control its chemical evolution, leading to the
formation of more complex (organic) molecules \citep{aea05}. Consequently, the CO chemistry
does not reach an equilibrium state even after 5~Myr of the
evolution in the disk midplane. The
characteristic timescale of turbulent diffusion in the disk is about $10^4$~years, and thus mixing
should modify the CO concentration mostly in the disk midplane.

\begin{figure}
\plotone{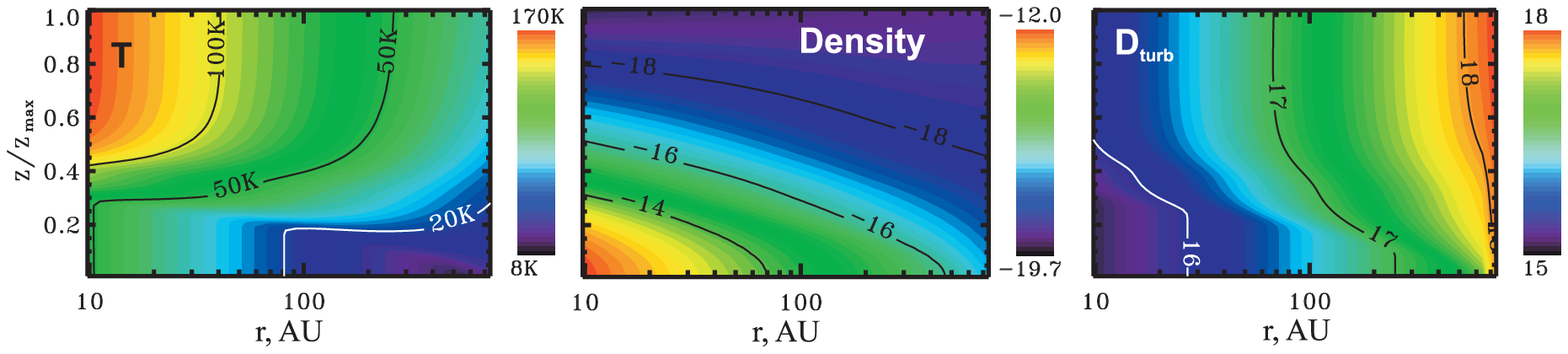}
\caption{Distributions of the temperature (left panel), mass density (middle panel, logarithmic scale), and the diffusion coefficient in cm$^2$\,s$^{-1}$
(right panel, logarithmic scale)
for the DM~Tau disk model. The Y-axis
represents the normalized vertical height, $z'(r) =
z(r)/z_{\max}(r)$. See the electronic edition
of the Journal for a color version of this figure.} \label{disk_struc}
\end{figure}

Indeed, the CO gas-phase abundances are substantially enhanced due
to turbulent diffusion in the outer midplane region, where $r\ga 80$~AU and
$z'\la 0.2$ (compare left and right panels in Fig.~\ref{abunds}).
This region is the coldest disk part where the temperature is mostly
below the freezing point of $\approx 20$\,K for CO, and can be as
low as 8\,K (Fig.~\ref{disk_struc}, left panel). Therefore, it
is not a surprise that in the static model the
final relative CO abundance for the outer midplane of the DM~Tau
disk does not exceed $\sim 10^{-10}$ and is typically
much lower.

Neither vertical, nor radial mixing alone can change the situation
in the DM~Tau disk model.
In contrast, 2D-mixing makes
the ``hole'' in the CO concentration less prominent in the coldest
region of the disk, and the entire zone of heavy CO depletion is
smaller than in the static case. The
average CO abundance is close to $\sim 10^{-10}$ and the upper
boundary of the strong CO depletion is located at a smaller height,
$z'\la 0.2$, compared to $z'$ of about 0.3 for the non-mixing
case. Changes in column densities through the cold midplane are
also significant. In Fig.~\ref{cold} (bottom panel) we show contributions to the CO column
density from the region where $T<25$\,K. At $r>200$\,AU, the model with only vertical
mixing shows lower CO column density compared to the static model due to more effective conversion CO to H$_2$CO on grain surfaces. In the model with radial mixing, the CO
column density is only enhanced at $r<200$\,AU, while in the 2D model it is higher than
in other models by a few orders of magnitude up to $r<400$\,AU, mainly due to outward transport of CO in radial direction.

What is the reason for the high midplane mixing efficiency in the 2D model?
It is usually assumed that the vertical mixing is more effective than
the radial mixing. This is correct for the disk as a whole. But close to
the midplane there is no vertical abundance gradient, so that vertical
mixing does not occur in this region. Therefore, vertical mixing is unable to remove the
midplane depletion region at around 100~AU, while being able to smear
out the elevated depletion area located at $r = 10-30$~AU (Fig. \ref{abunds},
second panel). The radial temperature (and abundance) gradient is steeper
in the midplane, so the mixing there occurs predominantly in radial
direction. In the model with only the radial mixing, gas-phase CO is
transported from the inner disk toward the outer midplane extending the
region of abundant CO from the initial radius of about 50~AU to more
than 100~AU and beyond.

The appearance of the maps in $r-z^\prime$ coordinates should not be misinterpreted. The
region of low CO abundance at $z^\prime\approx0.2$, which extends
over almost the entire disk, in fact, bends up at higher radii, following the disk
flaring. So, the linear distance that has to be traveled by species being mixed
is much greater in this region than closer to the midplane. Therefore, this region
cannot be destroyed by the radial mixing alone, and the vertical mixing is
needed to smoothen the CO distribution. The net outcome of this interplay
between temperature gradients, disk geometry, and chemistry is that both mixing
effects are needed to produce the high gas-phase abundance all over the
outer midplane region. The radial mixing brings gas-phase CO from
the warm inner part of the disk and from its outer part, where the freezing
timescale is somewhat longer, which causes somewhat higher midplane gas-phase CO abundances.
So, it is actually the radial mixing that causes enhanced gas-phase CO
in the very midplane. The vertical mixing lifts CO above the midplane
and helps to fill in gaps in the CO distribution, evident in 1D mixing models.
Thus, turbulent mixing is capable to
replenish the CO gas in the low temperature disk regions,
competing successfully with the opposite process of
CO freeze-out.

There seems to be an observational evidence toward this effect. Recently,
\citet{dartois} have found a reservoir of abundant CO
gas in the outer disk midplane. According to their data, this region of
nearly constant low temperature of $\sim 13$\,K is located at
$100~{\rm AU}\la r\la 650$~AU and $z < 0.2z_{\rm
max}$ (see also Fig.~\ref{disk_struc}, left panel). This is
the region where turbulent mixing tends
to increase the gas-phase CO concentration.

Even though the midplane CO abundance is increased by orders of magnitude in
the 2D model, it is still not high enough. Specifically, column density of cold $^{12}$CO at $T<20$\,K is $\la 10^{14}$~cm$^{-2}$ everywhere in the disk, which is too low to account for observations. In the 2D model (but not in other
models!) significant amount of CO is located in the region where the temperature is slightly
above the 20\,K threshold. If we set the limiting temperature to be 25\,K, the $^{12}$CO
column density raises to $10^{16}$~cm$^{-2}$ in the 2D model everywhere in the outer
part of the disk ($r>100$~AU), while in other models it reaches the similar level only at
 $r>500$~AU, still being less than $10^{13}$~cm$^{-2}$ at smaller radii (Fig.~\ref{cold}). The region with
$T<25$\,K contributes about 1/4 of the total LTE optical depth for $^{13}$CO in the 2D model. It must be noted that this temperature threshold is higher than the observed CO temperature of $\sim13$\,K \citep{dartois}. But in any case, this result clearly indicates
that further investigation of 2D (3D) mixing and advective transport with different disk models and/or chemical reaction sets is worthwhile.

Note that it is quite difficult to suggest other
robust non-thermal mechanism for such an enhancement in the outer,
cold and dark disk midplane, since free-floating CO molecules
should stick there to the grain surfaces with nearly $100\%$ probability
upon an encounter with a grain and never come back \citep{bea06}.

\citet{dea05} have proposed a chemical model in which icy mantles
are desorbed by the interstellar far UV radiation, forming a
layer of gas-phase water above the disk midplane. However, deeper
inside the disk the density increases so rapidly that the
penetration of even hard-energy FUV photons become next to
impossible. The interstellar UV radiation, penetrating the disk
from its boundaries, does not propagate deep enough,
either, to allow productive desorption of CO at $r\ga 100$~AU, so
only the outermost disk part around 700--800~AU is affected.

Another possible explanation would be substantial grain growth,
leading to a dramatic decrease of the disk shielding and thus rapid
CO mantle photodesorption. However, there is no observational evidence
(yet) for considerable grain growth in the case of DM~Tau. In addition,
we should note that an overall decrease of the disk opacity makes heating
by irradiation more efficient and thus decreases the mass reservoir of
the very cold gas at $T<20$\,K.

\acknowledgments
We thank P.~D'Alessio for providing the disk model. We are grateful to both referees
for their comments and suggestions that helped to improve the presentation drastically.
DS is supported by the {\it Deutsche
Forschungsgemeinschaft}, DFG project ``Research Group Laboratory
Astrophysics'' (He 1935/17-2). DW acknowledges support from the
RFBR grant 04-02-16637 and the RF President Grant
MD-4815.2006.2.


\clearpage

\clearpage

\clearpage

\end{document}